\documentstyle[12pt]{article}

\renewenvironment{thebibliography}[1]
{\normalsize
 \begin{list}{[\arabic{enumi}]}
 {\usecounter{enumi} \setlength{\parsep}{0pt}
  \setlength{\itemsep}{3pt} \settowidth{\labelwidth}{[#1]}
  \sloppy}}
{\end{list}}

\newcommand{\cleqn}{\setcounter{equation}{0}}

\newcommand{\pr}{\hspace{\parindent}}

\begin{document}

\hfill\vbox{\baselineskip14pt
            \hbox{\bf KEK-TH-420}
            \hbox{\bf KEK Preprint 94-164}
            \hbox{\bf December 1994}
            \hbox{H}}
\baselineskip20pt

\vspace*{1cm}
\begin{center}
{\Large{\bf Probing the Weak Boson Sector in $\gamma e\rightarrow Ze$}}
\end{center}
\vspace*{1cm}

\begin{center}
\large S.Y. Choi
\end{center}
\begin{center}
{Theory Group, KEK, Tsukuba, Ibaraki 305, Japan }
\end{center}
\vspace*{2cm}

\begin{center}
\large Abstract
\end{center}
\vspace{0.2cm}
\begin{center}
\begin{minipage}{14cm}
\baselineskip=20pt
\noindent
We study possible deviations from the standard model in the reaction
$\gamma e\rightarrow Ze$ at a 500 GeV $e^+e^-$ collider.
As a photon source we use a laser backscattered photon beam.
We investigate the most general $\gamma Z\gamma$ and $\gamma ZZ$
vertices including operators up to energy-dimension-six which
are Lorentz invariant. These vertices require four extra parameters;
two are CP-conserving, $h^\gamma_1$ and $h^Z_1$, and two are
CP-violating, $h^\gamma_2$ and $h^Z_2$.
We present analytical expressions of the helicity amplitudes
for the process $\gamma e\rightarrow Ze$ for
arbitrary values of anomalous couplings. Assuming Standard Model
values are actually measured we present the allowed region in the
($h^\gamma_1,h^Z_1$) plane at the 90\% confidence level.
We then show how the angular correlation of the $Z$ decay products
can be used to extract detailed information on the anomalous
(especially CP-violating) $\gamma Z\gamma$ and $\gamma ZZ$ couplings.
\end{minipage}
\end{center}
\vfill

\baselineskip=20pt
\normalsize

\newpage
\setcounter{page}{2}

\section{Introduction}
\cleqn

\pr
The Standard Model (SM) has very successfully explained
all available experimental data. However, even though
the properties of both the fermions and the vector bosons
are predicted  from a general gauge principle, only
the fermionic sector and its interaction structure with
gauge bosons have been rigorously tested
and verified. On the contrary, the non-Abelian gauge-group
structure involving the gauge-boson self couplings has yet to be
probed precisely.

To investigate the non-Abelian gauge-group structure of the
weak vector bosons, extensive studies\cite{Hagiwara1,Choi}
have been devoted to the anomalous $WWZ$ and $WW\gamma$ couplings.
But, little attention\cite{Renard,Ryzak,Barger,ZZ} has
been paid to studies of the anomalous couplings involving
only the neutral $\gamma$ and $Z$ bosons.
One reason could be that while the vertices $WWV$ ($V=\gamma,Z$)
already appear at the tree-level in the SM, $\gamma ZV$ vertices
are forbidden due to gauge symmetry. However, adopting a
general scheme to include all seven allowed anomalous trilinear
vector boson couplings\cite{Hagiwara1} in a
phenomenological analysis, we find that two anomalous $WWV$
couplings violate C invariance, and these C-violating terms
should be related with $ZZV$ and $\gamma ZV$ couplings by
the global SU(2) custodial symmetry\cite{Cust}.
{}From this point of view the possible trilinear couplings
in the neutral boson sector are an interesting topic at
high-energy hadron and $e^+e^-$ colliders.
Experimentally the processes involving neutral bosons (such as
$e^+e^-\rightarrow ZV$ and $\gamma e\rightarrow Ze$) allow for
high precision with a clean and easily analyzable final state.
Moreover, while the corresponding processes,
$e^+e^-\rightarrow W^+W^-$ and $\gamma e\rightarrow W\nu_e$,
involve many anomalous couplings, each of the $ZZ$ and $Z\gamma$
modes can have only two anomalous couplings (only one if CP
invariance is imposed). Therefore, we can identify unambiguously
the parameter from which a slight deviation originates.

Processes to which anomalous trilinear couplings of the neutral
bosons could contribute include radiative $Z$ decays,
$e^+e^-\rightarrow \gamma\gamma$, $Z\gamma$, $ZZ$,
and $\gamma e\rightarrow Ze$.
Until now the structure of the trilinear couplings of the neutral
bosons has been studied mainly in the radiative decays of $Z$
bosons\cite{Barger} and the processes, $e^+e^-\rightarrow Z\gamma$,
$ZZ$\cite{ZZ}. Rather recently the process $\gamma e\rightarrow Ze$
has been considered in Ref.\ \cite{Cornet}.
However, most studies of these anomalous neutral-boson
couplings were restricted to only a few couplings
which satisfy CP invariance and they have not fully employed
the angular correlations of the $Z$ decay products to extract
detailed information on the anomalous couplings.

In the present work we provide a more complete and systematic
study of the most general dimension-six
$\gamma Z\gamma$ and $\gamma ZZ$ vertices through the
process $\gamma e\rightarrow Ze$ (see Fig.~1)
with energetic laser-backscattered photons\cite{Ginzburg} at
a 500 GeV $e^+e^-$ linear collider (NLC).
We should note that the backscattered laser photon beam is
especially interesting for the investigation of the anomalous
$\gamma Z\gamma$ and $\gamma ZZ$ couplings in
$\gamma e\rightarrow Ze$; the spectrum of the scattered laser
photons is hard and the luminosity of the backscattered photon beam
can be as high as the luminosity of the original $e^\pm$ beam.
Furthermore, we can obtain from the initial laser beams,
whose polarization is easily controlled, highly polarized
backscattered $\gamma$ beams.

The paper is organized as follows. In Section~2 we give the most
general form of dimension-six $\gamma Z\gamma$ and $\gamma ZZ$ couplings
and show which constraints on these couplings come from electroweak
gauge symmetry, C and P. In Section~3 we present, in a compact form,
the complete helicity amplitudes, derived from the most general
couplings of Section~2, for the process $\gamma e\rightarrow Ze$.
Section~4 is devoted to the presentation of all nine coefficients
of the differential angular distributions of the $Z$
decay into a massless fermion-antifermion pair, expressed
in terms of the helicity amplitudes for the production process
$\gamma e\rightarrow Ze$. In Section~5 we introduce as a photon source
Compton backscattered laser light for which we discuss in detail
the effective luminosity.
Assuming the SM values of the differential cross section are
actually measured, we present, in Section~6, the allowed region of
the CP-conserving couplings, $h^\gamma_1$ and $h^Z_1$, at the 90\%
confidence level. Then we investigate the angular correlations of the
$Z$ boson decay products to extract detailed information on the
CP-violating anomalous couplings, $h^\gamma_2$ and $h^Z_2$.
In Section~7 we summarize our findings.

\section{Anomalous $\gamma Z\gamma$ and $\gamma ZZ$ couplings}
\cleqn

\pr
The most general, Lorentz invariant  effective
Lagrangian\cite{Hagiwara1} for $\gamma Z\gamma$ and $\gamma ZZ$
interactions, restricted to operators of dimension six, can be
written as
\begin{eqnarray}
&&{\cal L}_{\gamma Z}=\frac{g_{\gamma ZZ}}{m^2_Z}
  \left[\zeta_Z F_{\mu\nu}(\partial^\mu Z^\lambda\partial_\lambda Z^\nu
   -\partial^\nu Z^\lambda\partial_\lambda Z^\mu)+\eta_{1Z}
  (\partial_\mu Z^\alpha\partial_\alpha Z_\nu)\tilde{F}^{\mu\nu}
   +\eta_{2Z}(Z_\mu\partial^2 Z_\nu)\tilde{F}^{\mu\nu}\right]\nonumber\\
&&\hskip 1cm +\frac{g_{\gamma Z\gamma}}{m^2_Z}
   \left[\zeta_\gamma F_{\mu\nu}F^{\nu\lambda}(\partial^\mu Z_\lambda
      +\partial_\lambda Z^\mu)+\eta_{1\gamma}
       Z_\mu (\partial^\rho \tilde{F}^{\mu\nu}F_{\nu\rho})
      +\eta_{2\gamma} Z_\mu
      (\tilde{F}^{\mu\nu}\partial^\rho F_{\nu\rho})\right],
  \label{eff_L}
\end{eqnarray}
where $Z^\mu$ is the $Z$-boson field,
$F_{\mu\nu}=\partial_\mu A_\nu-\partial_\nu A_\mu$,
and
$\tilde{F}^{\mu\nu}
  =\frac{1}{2}\varepsilon^{\mu\nu\alpha\beta}F_{\alpha\beta}$.
Here we have neglected the scalar component of the $\gamma$ and $Z$ bosons:
\begin{eqnarray}
\partial^\mu A_\mu=0,\qquad  \partial^\mu Z_\mu=0.
\label{Scalar}
\end{eqnarray}
The condition (\ref{Scalar}) is automatic for on-shell $Z$ and $\gamma$.
It is also valid not only for the virtual photon but also for the $Z$ boson
in the process $\gamma e\rightarrow Ze$ in which
all terms with $\partial^\mu Z_\mu$
are in fact proportional to the electron mass and thus negligible.

We find that in momentum space the corresponding
$\gamma ZV$ vertex (for on-shell $Z$ and $\gamma$) shown in Fig.~2
can be expressed in the following simple form:
\begin{eqnarray}
\Gamma^{\alpha\beta\mu}_{\gamma ZV}(q_1,q_2,P)
       &=&i\left(\frac{P^2-m^2_V}{m^2_Z}\right)
   \left[h_1^V\varepsilon^{\mu\alpha\beta\rho}q_{1\rho}
       + h_2^V(q_1^\mu g^{\alpha\beta}-q_1^\beta g^{\mu\alpha})\right],
\end{eqnarray}
where $q_1$ and $q_2$ are four-momenta of the photon and the $Z$ boson,
respectively, and $P=q_1-q_2$.
Terms proportional to $P^\mu$ or $q_2^\beta$ are omitted because
they do not contribute for $m_e\rightarrow 0$.
Note that the above expression is manifestly electromagnetic gauge
invariant for the on-shell photon.
The coupling $h_1^V$ is CP-even and  $h_2^V$ is P-even: both are C-odd.
(See Table~1.) The overall factor $P^2-m^2_V$ comes from gauge
invariance for $V=\gamma$ and from Bose symmetry for $V=Z$.
This factor cancels the corresponding $t$-channel pole in the
process $\gamma e\rightarrow Ze$.

It is straightforward to calculate the contribution of the dim-6
operators (\ref{eff_L}) to the form factors $h^V_i$.
We obtain the relations
\begin{eqnarray}
&&h^\gamma_1=\zeta_\gamma,\qquad
  h^\gamma_2=\eta_{1\gamma}+\eta_{2\gamma},\nonumber\\
&&h^Z_1 =\zeta_Z,\qquad
  h^Z_2 = \frac{\eta_{1Z}}{2}+\eta_{2Z}.
\end{eqnarray}
The form factors, $h^V_i$, are real due to hermiticity and
the fact that the $t$-channel momentum transfer, $P^2$, is negative
in the process $\gamma e\rightarrow Ze$.

Without any loss of generality we can fix the overall coupling constants.
We choose, for convenience,
\begin{eqnarray}
g_{\gamma Z\gamma}=g_{\gamma ZZ}=e,
\end{eqnarray}
where $e$ denotes the positron charge, $s_W=\sin\theta_W$,
$c_W=\sqrt{1-s^2_W}$, and $\theta_W$ is the weak mixing angle
of the SM.

\section{Helicity amplitudes for $\gamma e\rightarrow Ze$}
\cleqn

\pr
In this section we present all the helicity amplitudes for the process
\begin{eqnarray}
\gamma(q_1,\lambda_1)+e^-(p_1,\sigma_1)\rightarrow
                     Z(q_2,\lambda_2)+e^-(p_2,\sigma_2),
\end{eqnarray}
as shown in Fig.~3, with the general $\gamma Z\gamma$ and $\gamma ZZ$
couplings. The four-momentum and the helicity of each particle are
shown in parentheses.

Clearly helicity amplitudes contain more information than the simple
cross section for a polarized $Z$ boson. The relative phases of
the amplitudes are important for the distribution of the final
fermions because the interference of different $Z$ helicity
states gives a nontrivial azimuthal-angle dependence.
It is straightforward to include polarization of
the initial $e$ and photon beams in the helicity amplitudes.

The helicity of a massive particle is not a relativistic invariant
but is frame-dependent. In this paper we define the helicity of
the $Z$ in the $\gamma e$ c.m. frame.
Let us present the results in a compact form using the helicity
basis for the $Z$.
For convenience we rewrite the amplitude by
extracting some kinematic factors as
\begin{eqnarray}
{\cal M}_{\sigma_1\lambda_1;\sigma_2\lambda_2}(\Theta)
       =
     eg_Z\delta_{\sigma_1\sigma_2}\left(1-1/r\right)^{1/2}
     d^{J_0}_{\Delta\lambda_1,\Delta\lambda_2}(\Theta)
     \tilde{{\cal M}}_{\sigma_1;\lambda_1\lambda_2}(\Theta),
\label{Ext}
\end{eqnarray}
where $r=s/m^2_Z$,
      $\Delta\lambda_1=\lambda_1-\frac{1}{2}\sigma_1$,
      $\Delta\lambda_2=\lambda_2-\frac{1}{2}\sigma_2$,
      $J_0={\rm max}(|\Delta\lambda_1|,|\Delta\lambda_2|)$,
and   $\Theta$ denotes the scattering angle of $Z$ boson
with respect to the photon direction in the $\gamma e$ c.m. frame.
The coupling constant $g_Z=e/s_Wc_W$.
The $d$ functions, $d^{J_0}_{\Delta\lambda_1,\Delta\lambda_2}(\Theta)$,
are given in the convention of Ref.\ \cite{Rose}.
The explicit form of the $d$ functions which are needed in
the present work is found in Table~2. Of course $\tilde{{\cal M}}$ in
eq.~(\ref{Ext}) is not a partial wave amplitude because
it may still be $\Theta$-dependent.

One of the three lowest-order diagrams, namely the one with the
$s$-channel exchange of an electron has only $J=\frac{1}{2}$
and $J=\frac{3}{2}$ partial waves due to conservation of angular
momentum conservation. On the other hand, the
diagrams with $t$-channel $\gamma$ or $Z$ exchange and $u$-channel
electron exchange may have all the partial waves with $J\geq J_0$.

We use the two-component spinor technique of Ref.\ \cite{Hagiwara2}
to calculate the helicity amplitudes. The results may be expressed as
a sum of $s$-, $u$-, and $t$-channel contributions as
\begin{eqnarray}
\tilde{\cal M}=\tilde{\cal M}_s+\tilde{\cal M}_u+\tilde{\cal M}_t,
\end{eqnarray}
where the individual contributions are given by
\begin{eqnarray}
\tilde{\cal M}_s&=&(v_e+\sigma a_e)A^{\lambda_1\lambda_2}_\sigma,\\
\tilde{\cal M}_u&=&-\frac{2(v_e+\sigma a_e)}{(1-1/r)(1+\cos\Theta)}
                   B^{\lambda_1\lambda_2}_\sigma,\\
\tilde{\cal M}_t&=&r\Bigl[\lambda_1 h_{1\sigma}-ih_{2\sigma}\Bigr]
                   C^{\lambda_1\lambda_2}_{\sigma},
\end{eqnarray}
with
\begin{eqnarray}
h_{i\sigma}=(v_e+\sigma a_e)h^Z_i-c_Ws_Wh^\gamma_i\qquad (i=1,2).
\end{eqnarray}
Here $v_e=s_W^2-1/4$ and $a_e=1/4$ in the SM.
The coefficients $A^{\lambda_1\lambda_2}_\sigma$ and
$B^{\lambda_1\lambda_2}_\sigma$ (which
arise purely from SM physics) are shown in Table~3.
The $C^{\lambda_1\lambda_2}_\sigma$ coefficients
(which come from for the anomalous $\gamma Z\gamma$
and $\gamma ZZ$ are shown in Table~4.

We first investigate the SM contributions.
$A^{\lambda_1\lambda_2}_\sigma$ and
$B^{\lambda_1\lambda_2}_\sigma$ are invariant under the
simultaneous reversal of all particle helicities as can be
seen from Table~3.
The couplings $(v_e+\sigma a_e)$ appear only as a
global factor in the SM matrix elements such that
the cross section for the right-handed electron is proportional to
that for the left-handed electron with an overall factor
$(v_e-a_e)^2/(v_e+a_e)^2\approx 0.7$.
We find that the amplitudes vanish for the electron and boson
polarizations: $(\sigma;\lambda_1,\lambda_2)=(\pm;\pm,\mp)$,
$(\pm;\pm,0)$.

For the electron and boson polarizations
the total cross sections are given by
\begin{eqnarray}
\sigma_{\rm tot}(|\cos\Theta|< 1-\epsilon:\sigma,\lambda_1,\lambda_2)
=3{\rm R}\left(\frac{v_e+\sigma a_e}{c_Ws_W}\right)^2
     T^{\lambda_1\lambda_2}_\sigma(\epsilon,r),
\label{SM_cs}
\end{eqnarray}
where
\begin{eqnarray}
&&{\rm R}=\frac{4\pi\alpha^2}{3s}=0.347
    \left[\frac{0.5{\rm TeV}}{\sqrt{s}}\right]^2 [{\rm pb}],\\
&&T^{++}_{+}(\epsilon,r)
  =T^{--}_{-}(\epsilon,r)=\left(1-\frac{1}{r}\right)^2
                    \log\left(\frac{2}{\epsilon}-1\right),\nonumber\\
&&T^{+-}_{+}(\epsilon,r)
  =T^{-+}_{-}(\epsilon,r)=0,\nonumber\\
&&T^{-+}_{+}(\epsilon,r)
  =T^{+-}_{-}(\epsilon,r)
  =\frac{1}{r^2}\left[\log\left(\frac{2}{\epsilon}-1\right)
  -\frac{3}{2}(1-\epsilon)\right],\nonumber\\
&&T^{--}_{+}(\epsilon,r)
  =T^{++}_{-}(\epsilon,r)
  =\frac{1}{2}\left(1-\epsilon\right),\nonumber\\
&&T^{+0}_{+}(\epsilon,r)
     =T^{-0}_{-}(\epsilon,r)=0,\nonumber\\
&&T^{-0}_{+}(\epsilon,r)
     =T^{+0}_{-}(\epsilon,r)
     =\frac{1}{r}\left(1-\epsilon\right).
\end{eqnarray}
The results (\ref{SM_cs}) have been compared to and are
consistent with those of Denner and Dittmaier\cite{Denner}.
We have introduced the small angular
cut $\epsilon$ ($|\cos\Theta|<1-\epsilon$) to regularize the
backward singularity caused by neglecting the small
electron mass; this also allows us to neglect particles lost
down the beam pipe.

We first note that, at threshold ($s=m^2_Z$ or $r=1$),
the $(\pm;\pm\pm)$ cross sections vanish while the others
remain finite except of course the two types which are
identically zero. Second, we find that, at high energies,
the dominant contributions are those where
the photon and the $Z$ boson have the same helicity.
Third, we find that  the cross section for longitudinal
$Z$ bosons is suppressed by at least $1/r$. Fig.~4 illustrates
these features of the SM cross sections for left-handed electrons
(integrated over $-0.9< \cos\Theta < 0.9$).
The angular dependence of the differential cross sections is
illustrated in Fig.~5. Clearly it is characterized by the
$u$-channel pole in the backward region and kinematical zeros
in the forward region.
The $(\pm;\mp,0)$ and $(\pm;\mp,\mp)$ cross
sections do not exhibit such a $u$-pole structure due to
its cancellation by a factor in the numerator of $u$ and $u^2$,
respectively.
On the other hand, the cross sections for $(\pm;\pm,\pm)$ and
$(\pm;\mp,\mp)$ are finite in the forward region
while those for $(\pm;\mp,0)$ and $(\pm;\pm,\mp)$ vanish as $t$
and as $t^2$, respectively.

Let us conclude this section with brief remarks on the consequences
of CP and CP$\tilde{\rm T}$ invariances\cite{Hagiwara1} on the
full amplitudes. Since the form factors $v_e$, $a_e$, and
$h^V_i$ for all $i$ are real, CP$\tilde{\rm T}$ invariance
leads to the relation:
\begin{eqnarray}
{\cal M}_{\sigma_1,\lambda_1;\sigma_2,\lambda_2}=
\bar{{\cal M}}^*_{-\sigma_1,-\lambda_1;-\sigma_2,-\lambda_2},
\label{CPT_amp}
\end{eqnarray}
up to an overall phase. (The overall phase is $-1$ in the
convention of Ref.\ \cite{Hagiwara2}.)
Here the amplitude $\bar{\cal M}$ is for
the process $\gamma e^+\rightarrow Ze^+$.
On the other hand, CP invariance leads to the relation
\begin{eqnarray}
{\cal M}_{\sigma_1,\lambda_1;\sigma_2,\lambda_2}=
\bar{{\cal M}}_{-\sigma_1,-\lambda_1;-\sigma_2,-\lambda_2}.
\label{CP_amp}
\end{eqnarray}
The relation (\ref{CP_amp}) can be directly used
as a test of CP conservation.

\section{Angular correlations of $Z$ boson decay products}
\cleqn

\pr
In this section we present the general angular distributions
of the decay products in the sequential process (See Fig.~6.)
\begin{eqnarray}
\gamma(q_1,\lambda_1)+e(p_1,\sigma_1)
     &\rightarrow& Z(q_2,\lambda_2)+e(p_2,\sigma_2),\nonumber\\
Z(q_2,\lambda_2)&\rightarrow&f(k_1,\rho_1)+\bar{f}(k_2,\rho_2).
\end{eqnarray}
The masses of the fermions are neglected.

Since the decay processes are well understood
the dependence of the cross section on the angles of the
final state fermions can be extracted explicitly. The fact
that the $Z$ boson has spin one allows maximally nine
angle-dependent terms\cite{Hagiwara3}. The coefficient of
each term can be written in terms of the production density
matrix for the $Z$ boson, which may be obtained from the polarization
amplitudes of Section~3.

Let us express the full amplitude as follows
\begin{eqnarray}
{\cal M}(p_1,\sigma_1;q_1,\lambda_1;p_2,\sigma_2;k_i,\rho_i)
  &=&D_Z(q^2_2)\sum_{\lambda_2}
{\cal M}(p_1,\sigma_1;q_1,\lambda_1;p_2,\sigma_2;q_2,\lambda_2)\nonumber\\
  &&\hskip 1.5cm \times {\cal M}_D(q_2,\lambda_2;k_1,\rho_1;k_2,\rho_2),
\end{eqnarray}
with the $Z$-boson propagator in the Breit-Wigner form:
\begin{eqnarray}
D_Z(q^2)=(q^2-m^2_Z+im_Z\Gamma_Z)^{-1}.
\end{eqnarray}
The production amplitude, ${\cal M}$, is the sum of the contributions
from three different channels as explicitly presented in Section~3.

In the limit of massless fermions the $Z$-boson decay amplitude,
${\cal M}_D$, for a given final-state fermion of helicity $\tau$
simplifies to
\begin{eqnarray}
{\cal M}_D(\lambda_2;\tau)
    =
 g_ZC\left(v_f+\tau a_f\right)
 2\sqrt{k^0_1k^0_2}X_D(\lambda_2;\tau).
  \label{Zdecay}
\end{eqnarray}
Here $v_f$ and $a_f$ are the standard vector and axial-vector couplings.
The effective color factor $C$ is 1 for leptons and $\sqrt{3}$ for
quarks. The $X_D$ can be explicitly derived in a given frame.

In the $\gamma e$ c.m. frame we choose the direction of the $Z$ momentum
as the positive $z$-axis and the direction of $\vec{q}_1\times \vec{q}_2$
as the $y$-axis; the scattering $\gamma e\rightarrow Z e$ takes place in
the $x$-$z$ plane. (See Fig.~7.) The production amplitude, ${\cal M}$,
is then a function of the scattering angle, $\Theta$,
as measured  between the momentum of the $\gamma$ and
the momentum of the $Z$ boson in this frame.
We express the decay amplitude, ${\cal M}_D$, in the rest frame
of the $Z$ boson, which is defined by a boost from the above
$\gamma e$ c.m. frame along the $z$-axis.

In the $Z$ rest frame we parametrize the four-moment of $f$ and
$\bar{f}$ as
\begin{eqnarray}
k^\mu_1&=&\frac{1}{2}\sqrt{q^2_2}
         (1,\sin\theta\cos\phi,\sin\theta\sin\phi,\cos\theta),\nonumber\\
k^\mu_2&=&\frac{1}{2}\sqrt{q^2_2}
         (1,-\sin\theta\cos\phi,-\sin\theta\sin\phi,-\cos\theta).
\end{eqnarray}
It is rather straightforward to evaluate $X_D$ of
eq.~(\ref{Zdecay}) in the $Z$ rest frame. We find
\begin{eqnarray}
&&X_D(\pm;+)=-\frac{1}{\sqrt{2}}(1\pm\cos\theta)e^{\pm i\phi},\nonumber\\
&&X_D(0;+)=X_D(0;-)=-\sin\theta,\nonumber\\
&&X_D(\pm;-)=\frac{1}{\sqrt{2}}(1\mp\cos\theta)e^{\pm i\phi}.
\end{eqnarray}

The polarization-summed squared matrix elements are now given by
\begin{eqnarray}
&&\sum|{\cal M}|^2=\sum_{\sigma_1}\sum_{\lambda_1}\sum_{\sigma_2}
  \sum_{\rho_1}\sum_{\rho_2}
  |{\cal M}(p_1,\sigma_1;q_1,\lambda_1;p_2,\sigma_2;k_i,\rho_i)|^2
   \nonumber\\
&&{ }\hskip 1.5cm = g^2_ZC^2q_2^2|D_Z(q_2^2)|^2\sum_{\lambda_1}\sum_\tau
  (v_f+\tau a_f)^2{\cal P}^{\lambda_1}_{\lambda_2\lambda^\prime_2}
  {\cal D}^\tau_{\lambda_2\lambda^\prime_2}.
\end{eqnarray}
Summation over repeated indices ($\lambda_2$ and $\lambda^\prime_2$) is
implied. The production tensor reads
\begin{eqnarray}
{\cal P}^{\lambda_1}_{\lambda_2\lambda^\prime_2}=\sum_{\sigma_1}
  \sum_{\sigma_2}{\cal M}_{\sigma_1,\lambda_1;\sigma_2,\lambda_2}(\Theta)
       {\cal M}^*_{\sigma_1,\lambda_1;\sigma_2,\lambda^\prime_2}(\Theta),
\end{eqnarray}
and the decay tensor reads
\begin{eqnarray}
{\cal D}^\tau_{\lambda_2\lambda^\prime_2}
        =X_D(\lambda_2;\tau)X^*_D(\lambda^\prime_2;\tau).
\end{eqnarray}

After integration over the square of the virtual $Z$ mass, $q^2_2$,
and summation over the final fermion polarization, $\tau$, the
differential cross section for a given photon helicity, $\lambda_1$,
can be expressed in the narrow $Z$ width approximation as
\begin{eqnarray}
\frac{{\rm d}\sigma(\lambda_1)}{{\rm d}\cos\Theta
      {\rm d}\cos\theta{\rm d}\phi}=\frac{(1-1/r)}{16\pi s}
    \frac{3{\rm B}(Z\rightarrow f\bar{f})}{16\pi}
    \sum_{i=1}^9F^{\lambda_1}_i{\cal D}_i(\theta,\phi).
\end{eqnarray}
Here the ${\cal D}_i$'s are the following nine orthogonal
functions which are normalized to $4\pi$:
\begin{eqnarray}
&&{\cal D}_1=1,\nonumber\\
&&{\cal D}_2=\frac{\sqrt{5}}{2}(1-3\cos^2\theta), \nonumber\\
&&{\cal D}_3=\sqrt{3}\cos\theta,\nonumber\\
&&{\cal D}_4=\sqrt{3}\sin\theta\cos\phi,\nonumber\\
&&{\cal D}_5=\frac{\sqrt{15}}{2}\sin(2\theta)\cos\phi,\nonumber\\
&&{\cal D}_6=\frac{\sqrt{15}}{2}\sin^2\theta\cos(2\phi),\nonumber\\
&&{\cal D}_7=\sqrt{3}\sin\theta\sin\phi,\nonumber\\
&&{\cal D}_8=\frac{\sqrt{15}}{2}\sin(2\theta)\sin\phi\nonumber\\
&&{\cal D}_9=\frac{\sqrt{15}}{2}\sin^2\theta\sin(2\phi).
\label{D-function}
\end{eqnarray}
Note that after integration over the decay angles $\theta$ and $\phi$
only the coefficient $F^{\lambda_1}_1$ remains.
The coefficients $F^{\lambda_1}_i$ are expressed in
terms of ${\cal P}^{\lambda_1}_{\lambda_2\lambda^\prime_2}$ as
\begin{eqnarray}
&&F^{\lambda_1}_1=\frac{2}{3}\left[{\cal P}^{\lambda_1}_{++}
                 +{\cal P}^{\lambda_1}_{00}
                 +{\cal P}^{\lambda_1}_{--}\right],\nonumber\\
&&F^{\lambda_1}_2=\frac{\sqrt{5}}{15}
      \left[2{\cal P}^{\lambda_1}_{00}-{\cal P}^{\lambda_1}_{++}-
             {\cal P}^{\lambda_1}_{--}\right],\nonumber\\
&&F^{\lambda_1}_3=\frac{1}{\sqrt{3}}
      \left(\frac{2v_fa_f}{v^2_f+a^2_f}\right)
      \left[{\cal P}^{\lambda_1}_{++}
           -{\cal P}^{\lambda_1}_{--}\right],\nonumber\\
&&F^{\lambda_1}_4=\frac{\sqrt{2}}{\sqrt{3}}{\rm Re}
      \left[{\cal P}^{\lambda_1}_{+0}
           -{\cal P}^{\lambda_1}_{-0}\right],\nonumber\\
&&F^{\lambda_1}_5=\frac{\sqrt{2}}{\sqrt{15}}
    \left(\frac{2v_fa_f}{v^2_f+a^2_f}\right)
    {\rm Re}\left[{\cal P}^{\lambda_1}_{+0}
           +{\cal P}^{\lambda_1}_{-0}\right],\nonumber\\
&&F^{\lambda_1}_6=-\frac{2}{\sqrt{15}}{\rm Re}
    \left[{\cal P}^{\lambda_1}_{+-}\right], \nonumber\\
&&F^{\lambda_1}_7=\frac{\sqrt{2}}{\sqrt{3}}
    {\rm Im}\left[{\cal P}^{\lambda_1}_{+0}
           -{\cal P}^{\lambda_1}_{-0}\right],\nonumber\\
&&F^{\lambda_1}_8=\frac{\sqrt{2}}{\sqrt{15}}
    \left(\frac{2v_fa_f}{v^2_f+a^2_f}\right)
    {\rm Im}\left[{\cal P}^{\lambda_1}_{+0}
          +{\cal P}^{\lambda_1}_{-0}\right],\nonumber\\
&&F^{\lambda_1}_9=\frac{2}{\sqrt{15}}
    {\rm Im}\left[{\cal P}^{\lambda_1}_{+-}\right].
\label{F_ftn}
\end{eqnarray}

Let us first examine the consequences of CP invariance (\ref{CP_amp})
on the nine coefficients (\ref{F_ftn}). Since the relevant
initial $\gamma e^-$ state is not CP-invariant, the CP transformation
should relate the process $\gamma e^-\rightarrow Ze^-$
to the CP-conjugate process $\gamma e^+\rightarrow Ze^+$.
Then CP invariance leads to the following relation in the
two full production and decay angular distributions:
\begin{eqnarray}
{\rm d}\sigma(\lambda_1;\Theta;\theta,\phi)\stackrel{\rm CP}{=}
{\rm d}\bar{\sigma}(-\lambda_1;\Theta;\pi-\theta,\pi-\phi),
\label{CP_rel}
\end{eqnarray}
where ${\rm d}\sigma({\rm d}\bar{\sigma})$ is the differential
cross section for the process
$\gamma e^-\rightarrow Ze^-(\gamma e^+\rightarrow Ze^+)$
including the decay of the $Z$.

We next examine the implications of having no absorptive
part in the amplitude. We find that CP$\tilde{\rm T}$
invariance (\ref{CPT_amp}) leads to the relation
\begin{eqnarray}
{\cal P}^{\lambda_1}_{\lambda_2\lambda^\prime_2}=
\bar{{\cal P}}^{-\lambda_1}_{-\lambda^\prime_2,-\lambda_2},
\label{CPT_rel1}
\end{eqnarray}
and for the full decay angular distributions it leads to the relation
\begin{eqnarray}
{\rm d}\sigma(\lambda_1;\Theta;\theta,\phi)
  \stackrel{{\rm CP}\tilde{\rm T}}{=}
{\rm d}\bar{\sigma}(-\lambda_1;\Theta;\pi-\theta,\pi+\phi).
\label{CPT_rel2}
\end{eqnarray}

The CP and CP$\tilde{\rm T}$ properties of the eighteen
coefficients, $F_i^\lambda$ and $\bar{F}^\lambda_i$ for the
CP-conjugate process $(i=1$ to $9$), are
listed in Table~5. It is straightforward to check that
the relation (\ref{CPT_rel1}) forces every CP$\tilde{\rm T}$-odd
coefficient to vanish. We emphasize once more that this is due to
the negative $t$-channel momentum transfer and the hermiticity
of the Lagrangian. As a result only nine coefficients among
the eighteen original coefficients survive.
Among these are six CP-conserving coefficients and three which are
CP-violating.

\section{Photon spectra}
\cleqn

\pr
The calculations of the previous sections were carried out for
monochromatic photons. In this section we consider
realistic photon colliders with the inevitable photon energy spread.
The subprocess $\gamma e\rightarrow Ze$ is then related to
$e^+e^-$ collisions by folding the cross section with an
appropriate differential $e\gamma$ luminosity function,
${\cal L}_{e\gamma}(\hat{s})$;
\begin{eqnarray}
\sigma
 =\int^s_0(d\hat{s}/\hat{s}){\cal L}_{e\gamma}(\hat{s})\sigma(\hat{s}).
\end{eqnarray}
Here $\hat{s}$ is the squared c.m. energy of the $\gamma e$
system.
We may consider three different photon sources; classical
bremsstrahlung, beamstahlung\cite{Blan} and the Compton backscattered
laser beam\cite{Ginzburg}.
Of these three the laser backscattered beam is most interesting in
our context since (1) the energy spectrum of the resulting photon
beam is very hard compared to the standard bremsstrahlung photons
and a typical beamstrahlung photons, (2) the effective luminosity
remains as high as the original $e^\pm$ beam, and (3) highly
polarized backscattered $\gamma$ beams are naturally produced
from polarized laser beams.
In light of these distinct features, we will confine ourselves to
the laser backscattered photon beam for the actual numerical analysis.

To describe the machine parameters in the laser backscattering process
we introduce the dimensionless variables
\begin{eqnarray}
x_0=\frac{4E\omega_0}{m^2_e},\qquad  x=\frac{\omega}{E},
\end{eqnarray}
where $E$ is the electron or positron beam energy
$\omega_0$ is the energy of the laser photon and $\omega$ is
the energy of the scattered photon. In this work $E=250$ GeV.
The maximum energy fraction of the scattered photon is given by
\begin{eqnarray}
x_m=\frac{x_0}{x_0+1}.
\end{eqnarray}
The value of $x_0$ value should be less than $2+2\sqrt{2}$ to
prevent a significant drop of conversion efficiency due to the
onset\cite{Ginzburg} of $e^+e^-$ pair production between backscattered
photons and laser photons. We take $x_0=2+2\sqrt{2}$, which is
the maximally allowed value. This means that for the 250 GeV electrons
or positrons we use an initial laser beam with
$\omega_0\approx 1.26$ eV.

The effective luminosity is very sensitive to the product
of the electron helicity, $\sigma$, and the laser photon helicity,
$\lambda_\gamma$. A more negative average vale of
$\sigma\cdot\lambda_\gamma$ gives
a harder and more monochromatic photon spectrum.
However, from the experimental point of view, the introduction
of polarized electron beams may lead to new systematic errors.
Additionally, the laser photons should collide with positrons, which
are at present difficult to polarize, in order to use
the polarized electron beam in a realistic $e^+e^-$ collider.
Therefore we assume that the electron and positron
beams are unpolarized. On the other hand, the laser can be easily
and completely polarized, and this polarization can serve as an
important experimental tool. We assume that the laser is completely
polarized $|\lambda_\gamma|=1$.
The final photon polarization depends on the electron or positron
beam energy, and it is proportional to the initial laser beam helicity.
The average helicity of the scattered laser photons is then given by
\begin{eqnarray}
\xi_2
 =
 -\lambda_\gamma\frac{(2r^\prime-1)(2-2x+x^2)}{2-2x+x^2-4r(1-r)(1-x)},
\end{eqnarray}
with the definition $r^\prime=x/x_0(1-x)<1$.
The $x$-dependence of the average photon helicity is illustrated
in  Fig.~8.
The effective photon luminosity is then given by
\begin{eqnarray}
&&{\cal L}_{e\gamma}(\hat{s})={\cal L}_{e\gamma}(xs)\nonumber\\
&&\hskip 1.5cm
  =\frac{x_0^2[2-2x+x^2-4r^\prime(1-r^\prime)(1-x)]}{(x_0^2-4x_0-8)
   \log(1+x_0)+x_0^2/2+8x_0-x_0^2/2(1+x_0^2)^2}.
\end{eqnarray}
Fig.~9 shows the $x$ dependence of the effective luminosity and
its components according to the final photon helicity.

\section{Discovery limits}
\cleqn

\pr
{}From the discussion of Section~3 it is apparent that different
anomalous $\gamma Z\gamma$ and $\gamma ZZ$ lead to deviations
of different helicity amplitudes from their SM values.
In order to discover and then distinguish the anomalous couplings from
each other, we thus have to separate the various helicity amplitudes.
As has been discussed in Section~4 the unique way to do this is
to study angular distributions of the $Z$ decay products.

The complete expression for the angular distribution of the
fermion-antifermion pair arising from the decay of the $Z$ boson
was given in section~4.
These angular distributions are particularly simple when
measured in the rest frame of the parent $Z$. Experimentally
this will require the direction of the $Z$-boson momentum
to be reconstructed to determine $\Theta$, the angle of the $Z$
with respect to the photon beam. The momenta of the decay
products (two jets or a charged lepton pair, say) can then be boosted
to the rest frame of their parent, which is moving with the velocity
$\beta_Z=(s-m^2_Z)/(s+m^2_Z)$ along the $Z$-boson axis.

We now cast the differential cross sections for
the processes $\gamma e^\pm\rightarrow Ze^\pm$ into the following form:
\begin{eqnarray}
&&{\rm d}\sigma(\lambda_\gamma)
       \sim\sum^9_{i=1}{\cal F}_i(\lambda_\gamma;\Theta;s)
       {\cal D}_i(\theta,\phi),\nonumber\\
&&{\rm d}\bar{\sigma}(\lambda_\gamma)
       \sim\sum^9_{i=1}\bar{{\cal F}}_i(\lambda_\gamma;\Theta;s)
       {\cal D}_i(\theta,\phi),
\end{eqnarray}
where the coefficients ${\cal F}_i$'s and $\bar{\cal F}_i$'s are
given by
\begin{eqnarray}
&&{\cal F}_i(\lambda_\gamma;\Theta;s)
           =\int_0^1dx {\cal L}_{e\gamma}(xs)
            \left[(1+\xi_2)F^+_i(\Theta;xs)
                 +(1-\xi_2)F^-_i(\Theta;xs)\right],\nonumber\\
&&\bar{{\cal F}}_i(\lambda_\gamma;\Theta;s)
           =\int_0^1dx {\cal L}_{e\gamma}(xs)
            \left[(1+\xi_2)\bar{F}^+_i(\Theta;xs)
                 +(1-\xi_2)\bar{F}^-_i(\Theta;xs)\right].
\end{eqnarray}
While the functions ${\cal D}_i$ reflect the known dynamics of the
$Z$ decay, the factors ${\cal F}_i$ and $\bar{\cal F}_i$ contain
information on the dynamics of the production processes,
$\gamma e^\pm\rightarrow Ze^\pm$. One can use the nine orthogonal
functions, ${\cal D}_i$, to extract much information on the production
mechanism.

First, we discuss CP-even distributions which are
in general, in the SM, non-vanishing at the tree level.
In the process $\gamma e\rightarrow Ze$ we have two parameters,
$h^\gamma_1$ and $h^Z_1$, which determine the size of
new CP-conserving contributions.

The simplest CP-even distribution is the differential cross
section ${\rm d}\sigma/{\rm d}\cos\Theta$, which is shown in Fig.~10
for the SM ($h^\gamma_1=h^Z_1=0$), for $h^\gamma_1=0.1$ and
for $h^Z_1=0.1$ at $\sqrt{s}=500$ GeV. It is clear that the differential
cross section is more sensitive to the coupling $h^\gamma_1$ than
to the coupling $h^Z_1$.  To make a quantitative estimate of
the sensitivity of the differential cross section to the CP-conserving
couplings we perform a $\chi^2$ analysis by comparing the SM
predictions with those corresponding to
non-vanishing anomalous couplings $h^\gamma_1$ and $h^Z_1$.
We compute the statistical errors from the following set of
NLC parameters:
(1) $\sqrt{s}=0.5$ TeV, $\int{\cal L}_{e^+e^-}=10{\rm fb}^{-1},$
    $|\cos\Theta|<0.9$,
(2) $Z$ reconstruction efficiency (including branching ratios) $=0.5$,
    and
(3) the SM cross section of the process $\gamma e\rightarrow Ze$
    as measured.
The systematic uncertainty is taken to be $\pm 5\%$.
Then we can derive bounds on the anomalous
couplings ($h^\gamma_1,h^Z_1$) at the 90\% confidence level
($\chi^2=4.61$) and display the contours in the
$(h^\gamma_1,h^Z_1)$ plane.
Fig.~11 shows the allowed regions for the anomalous couplings
$h^\gamma_1$ and $h^Z_1$ as determined from the study of
$\gamma e\rightarrow Ze$ using backscattered laser beams for
the (a) right-handed and (b) left-handed
initial laser beams, respectively, at $\sqrt{s}=0.5$ TeV.
We find that the constraint on the coupling $h^Z_1$ is dependent
a little on the laser beam helicity, but the constraint on
the coupling $h^\gamma_1$ is almost independent
of the helicity. From Fig.~11 we conclude that with the backscattered
laser photon beam we obtain the following constraints:
\begin{eqnarray}
-0.06<h^\gamma_1<0.04,\qquad -0.08<h^Z_1<0.08.
\end{eqnarray}

Next, we consider the CP-violating terms proportional to
$h^\gamma_2$ and $h^Z_2$.
Because of no absorptive parts the CP-violating terms contribute
solely to imaginary parts of the helicity amplitudes and they
would be not so large. Furthermore, the SM amplitudes are real
so that the CP-violating terms have very little effect
on real distributions such as ${\rm d}\sigma/{\rm d}\cos\Theta$.
A large sensitivity can be obtained only by measuring coefficients
of sines of azimuthal angles, where  the relative phases between
different helicity amplitudes interfere with the imaginary part due
to the CP-violating terms.

We have shown in Section~4 that three CP-violating
distributions can be considered in the processes
$\gamma e^\pm\rightarrow Ze^\pm$. Two distributions,
$F^{\lambda_1}_7-\bar{F}^{-\lambda_1}_7$ and
$F^{\lambda_1}_8-\bar{F}^{-\lambda_1}_8 $, require the interference
of the longitudinally polarized $Z$ boson and the transversely
polarized $Z$ boson, while the other distribution
$F^{\lambda_1}_9+\bar{F}^{-\lambda_1}_9$ involves only transversely
polarized $Z$ bosons. From Table~4 it is clear that the first two
distributions become more sensitive than the latter distribution as
the c.m. energy increases.
On the other hand, the distribution
$F^{\lambda_1}_8-\bar{F}^{-\lambda_1}_8$
requires the measurement of the final fermion polarization which is
possible only in the $\tau$-lepton decay mode of the $Z$ boson.
Without the actual polarization measurement, the effectiveness of
this distribution is reduced by a polarization
factor of $2v_fa_f/(v^2_f+a^2_f)$ as can be seen clearly in
eq.~(\ref{F_ftn}). Besides, charge identification
is required to determine the azimuthal angle of the final fermion.
Even if this requirement can be fulfilled in the charged-lepton mode
of the $Z$ decay, the leptonic $Z$ decay rates are
rather small and their polarization factor, $2v_fa_f/(v^2_f+a^2_f)$,
is very small (about $-0.08$).
Consequently after folding the effective photon luminosity
function with the $\gamma e\rightarrow Ze$ cross section we find that
the most sensitive measure of CP-violation is given by the following
asymmetry:
\begin{eqnarray}
A_{CP}(\lambda_\gamma)
  =\frac{{\cal F}_7(\lambda_\gamma)
   -\bar{\cal F}_7(-\lambda_\gamma)}{{\cal F}_1(\lambda_\gamma)
   +\bar{\cal F}_1(-\lambda_\gamma)}.
\label{CP-asymmetry}
\end{eqnarray}
The distribution ${\cal F}_7$, which is the coefficient of
$\sin\theta\sin\phi$, essentially denotes the up-down asymmetry
of the final fermion with respect to the scattering plane.

Fig.~12 shows the $\Theta$-dependence of the CP-odd asymmetry
(\ref{CP-asymmetry}) for $\lambda_\gamma=1$ and $h^\gamma_2=0.01$
(solid line), for $\lambda_\gamma=-1$ and $h^\gamma_2=0.01$
(dotted line), for $\lambda_\gamma=1$ and $h^Z=0.02$
(long-dashed line), and for $\lambda_\gamma=-1$ and $h^Z_2=0.02$,
respectively, at $\sqrt{s}=0.5$ TeV. A few interesting features
for the CP-odd asymmetry $A_{CP}(\lambda_\gamma)$ are noted. First,
the asymmetry due to the coupling $h^\gamma_2$ is very sensitive
to the scattering angle, $\Theta$, while the asymmetry due to
the $h^Z_2$ is rather insensitive to this angle.
Second, the asymmetry due to $h^\gamma_2$ is almost independent
of the laser photon helicity, but the asymmetry due to $h^Z_2$ is
strongly dependent on the laser photon helicity.
Third, in the backward region the asymmetry is more sensitive
to $h^\gamma_2$, but in the forward region the asymmetry
becomes very small in both cases. Since the differential cross section
is peaked by the $u$-channel pole, the CP-odd
asymmetry is much more sensitive to the coupling $h^\gamma_2$ than
the coupling $h^Z_2$. Consequently a deviation from the identically
zero prediction of the SM is clearly visible for the
$h^\gamma_2$ and the $h^Z_2$ of order of $10^{-2}$ with a left-handed
initial laser beam at the backward scattering region.

\section{Summary}
\cleqn

\pr
In this paper we have systematically studied observable
experimental distributions in the process
$\gamma e^\pm\rightarrow Ze^\pm$ with photons generated by
backward Compton-scattered laser light.
The process could serve as a probe of possible anomalous
$\gamma Z\gamma$ and $\gamma ZZ$ couplings along with the process
$e^+e^-\rightarrow Z\gamma$. Since the $Z$'s decay into
fermion-antifermion pairs, one may use the angular distributions
of the $Z$ decay products as polarimeters to efficiently analyze
the helicities of produced $Z$ bosons. Because the $Z$ decay
properties are well known, a careful study of the reaction
$\gamma e\rightarrow Ze\rightarrow f\bar{f}e$ therefore reveals
detailed information on anomalous $\gamma Z\gamma$ and $\gamma ZZ$
couplings through the angular correlations of the final-state
fermions.

More specifically we have shown that at an NLC
with the c.m energy $0.5$ TeV a search for anomalous moments
connected with the $\gamma Z\gamma$ and $\gamma ZZ$ vertices
is feasible. We have presented the allowed region for the
CP-conserving anomalous couplings, $h^\gamma_1$ and $h^Z_1$,
at the 90\% confidence level from the measurement of the
differential angular distributions for the production process
$\gamma e\rightarrow Ze$.

The effects of the CP-violating anomalous couplings
would not be particularly visible in the $Z$ angular distributions
in the production process. However a careful study of the polar and
azimuthal distributions of final-state leptons and anti-leptons
and a good use of laser beam polarization
enable us to isolate these CP-violating effects and to separate
the contribution of $h^\gamma_2$ from the contribution of $h^Z_2$.
We found that the most sensitive CP-violation asymmetry is the
up-down asymmetry of the final-state fermion with the respect to the
$\gamma e$ scattering plane.

We considered the sequential processes
$\gamma e\rightarrow Ze$, $Z\rightarrow f\bar{f}$ only to the
lowest order in electroweak interactions. It is certain that
a detailed study of possible anomalous contributions should
be extended to electroweak radiative corrections.
In our considerations we have treated the production
process separate and the decay processes separately,
and primarily studied kinematical effects; it should be relatively
straightforward to include radiative corrections separately for
the production and decay processes.
As a matter of fact, the electroweak radiative
corrections\cite{Denner} for the production process,
$\gamma e\rightarrow Ze$, have been done in Ref.\ \cite{Denner}.
One also can find an extensive
literature on the electroweak radiative corrections for
the $Z$ decay processes. Even though these corrections
should modify our amplitudes in detail, we expect that
the corrections do not change the overall structure of the
amplitudes.

To conclude, we found that at the NLC with a c.m. energy $0.5$ TeV
, an integrated luminosity $10{\rm fb}^{-1}$ and a polarized beam of
backscattered photons one may obtain the bounds $-0.06<h^\gamma_1<0.04$
and $-0.08<h^Z_1<0.08$ at the 90\% confidence level.
The CP-violating couplings $h^\gamma_2$ and $h^Z_2$ of the order
of $10^{-2}$ may easily be identified through the measurement of
the CP-violation up-down asymmetry.

\section*{Acknowledgements}

The author would like to thank the Japanese Ministry of Education,
Science and Culture for the award of a visiting fellowship.
He also would like to thank R.~Szalapski for careful reading
the manuscript and useful discussions.

\newpage

\newcommand{\prd}[1]{Phys.~Rev.~D{#1}}
\newcommand{\plb}[1]{Phys.~Lett.~B{#1}}
\newcommand{\npb}[1]{Nucl.~Phys.~B{#1}}
\newcommand{\zpc}[1]{Z.~Phys.~C{#1}}

\section*{References}

\newpage

\section*{Tables}
\renewcommand{\labelenumi}{\bf Table {\arabic {enumi}} \\}
\begin{enumerate}

\vspace*{1cm}
\item{Properties of the couplings $h^V_i$ $(V=\gamma,Z)$ under
      discrete transformations}

\vspace*{0.5cm}
\begin{tabular}{l|cc}\hline
{ } \\
$i$\hskip 1cm &\hskip 1cm $1$\hskip 1cm &\hskip 1cm $2$\hskip 1cm \\
{ } \\
\hline
{ } \\
P\hskip 1cm &\hskip 1cm $-$\hskip 1cm &\hskip 1cm $+$\hskip 1cm \\
{ } \\
CP\hskip 1cm &\hskip 1cm $+$\hskip 1cm &\hskip 1cm $-$\hskip 1cm \\
{ } \\
C\hskip 1cm &\hskip 1cm $-$\hskip 1cm &\hskip 1cm $-$\hskip 1cm  \\
{ } \\
\hline
\end{tabular}

\vspace*{1cm}
\item{Explicit form of the $d$ functions needed}

\vspace*{0.5cm}

\begin{tabular}{l}\hline
{ }   \\
$d^{3/2}_{3/2,3/2}(\Theta)=d^{3/2}_{-3/2,-3/2}(\Theta)
               =\frac{1}{2}(1+\cos\Theta)\cos\frac{\Theta}{2}$\\
{ }  \\
$d^{3/2}_{3/2,1/2}(\Theta)=-d^{3/2}_{-3/2,-1/2}(\Theta)
               =\frac{\sqrt{3}}{2}(1+\cos\Theta)\sin\frac{\Theta}{2}$\\
{ }  \\
$d^{3/2}_{3/2,-1/2}(\Theta)=d^{3/2}_{-3/2,1/2}(\Theta)
               =\frac{\sqrt{3}}{2}(1-\cos\Theta)\cos\frac{\Theta}{2}$\\
{ }  \\
$d^{3/2}_{1/2,-3/2}(\Theta)=d^{3/2}_{-1/2,3/2}(\Theta)
               =\frac{\sqrt{3}}{2}(1-\cos\Theta)\cos\frac{\Theta}{2}$\\
{ }  \\
$d^{1/2}_{1/2,1/2}(\Theta)=d^{1/2}_{-1/2,-1/2}(\Theta)
               =\cos\frac{\Theta}{2}$\\
{ }  \\
$d^{1/2}_{1/2,-1/2}(\Theta)=-d^{1/2}_{-1/2,1/2}(\Theta)
               =-\sin\frac{\Theta}{2}$\\
{ }  \\
\hline
\end{tabular}

\vspace*{1cm}
\item{Coefficients $A^{\lambda_1\lambda_2}_\sigma$ and
       $B^{\lambda_1\lambda_2}_\sigma$ for the standard model}

\vspace*{0.5cm}
\begin{tabular}{c|cc} \hline
{ }  \\
$(\sigma;\lambda_1,\lambda_2)$\hskip 0.5cm &
       \hskip 0.5cm $A^{\lambda_1\lambda_2}_\sigma$ &
        $B^{\lambda_1\lambda_2}_\sigma$\\
{ }  \\
\hline
{ }  \\
$(+;++),\ \ (-;--)$\hskip 0.5cm &\hskip 0.5cm $-2$
                  & $(1-1/r)(1-\cos\Theta)$   \\
{ } \\
$(+;+-),\ \ (-;-+)$\hskip 0.5cm &\hskip 0.5cm $0 $ & $ 0 $ \\
{ } \\
$(+;-+),\ \ (-;+-)$\hskip 0.5cm &\hskip 0.5cm $0 $
                  & $ 2\sqrt{3}/(3r)$    \\
{ } \\
$(+;--),\ \ (-;++)$\hskip 0.5cm &\hskip 0.5cm $0 $ & $ 2 $ \\
{ } \\
$(+;+0),\ \ (-;-0)$\hskip 0.5cm &\hskip 0.5cm $-\sqrt{2r}$ &
       $-\sqrt{\frac{r}{2}}(1-1/r)(1+\cos\Theta)$ \\
{ } \\
$(+;-0),\ \ (-;+0)$\hskip 0.5cm &\hskip 0.5cm $0 $
                  & $2\sqrt{2}/\sqrt{3r}$\\
{ }  \\
\hline
\end{tabular}

\vspace*{1cm}
\item{Coefficients $C^{\lambda_1\lambda_2}_{\sigma}$ for
      the general coupling}

\vspace*{0.5cm}
\begin{tabular}{c|c} \hline
{ }  \\
$(\sigma;\lambda_1,\lambda_2)$\hskip 0.5cm &\hskip 0.5cm
        $C^{\lambda_1\lambda_2}_{2\sigma}$\\
{ }  \\
\hline
{ }  \\
$(+;++),\ \ (-;--)$\hskip 0.5cm &\hskip 0.5cm $1 $     \\
{ }  \\
$(+;+-),\ \ (-;-+)$\hskip 0.5cm &\hskip 0.5cm $0 $      \\
{ }  \\
$(+;-+),\ \ (-;+-)$\hskip 0.5cm &\hskip 0.5cm $1/\sqrt{3}$ \\
{ }  \\
$(+;--),\ \ (-;++)$\hskip 0.5cm &\hskip 0.5cm $1 $       \\
{ }  \\
$(+;+0),\ \ (-;-0)$\hskip 0.5cm &\hskip 0.5cm $\sqrt{r/2}$ \\
{ }  \\
$(+;-0),\ \ (-;+0)$\hskip 0.5cm &\hskip 0.5cm $(r+1)/\sqrt{6r}$ \\
{ }  \\
\hline
\end{tabular}

\vspace*{1cm}
\item{CP and CP$\tilde{\rm T}$ properties of the 18 coefficients
      $F^{\lambda_1}$'s and $\bar{F}^{\lambda_1}$'s.}

\vspace*{0.5cm}
\begin{tabular}{cccc} \hline
{ }  \\
CP \hskip 0.3cm &\hskip 0.5cm  CP$\tilde{\rm T}$\hskip 0.5cm &
     \hskip 0.5cm angular coefficients
     \hskip 0.5cm & Number \\
{ }  \\
\hline
{ }  \\
even \hskip 0.3cm &\hskip 0.5cm even \hskip 0.5cm &
     $F^{\lambda_1}_1+\bar{F}^{-\lambda_1}_1$,
     $F^{\lambda_1}_2+\bar{F}^{-\lambda_1}_2$,
     $F^{\lambda_1}_3+\bar{F}^{-\lambda_1}_3$
     & 6  \\
{ }  \\
{ } \hskip 0.3cm &\hskip 0.5cm { } \hskip 0.5cm  &
     $F^{\lambda_1}_4-\bar{F}^{-\lambda_1}_4$,
     $F^{\lambda_1}_5+\bar{F}^{-\lambda_1}_5$,
     $F^{\lambda_1}_6+\bar{F}^{-\lambda_1}_6$
    & { }  \\
{ }  \\
even \hskip 0.3cm &\hskip 0.5cm odd\hskip 0.5cm &
     $F^{\lambda_1}_7+\bar{F}^{-\lambda_1}_7$,
     $F^{\lambda_1}_8+\bar{F}^{-\lambda_1}_8$,
     $F^{\lambda_1}_9-\bar{F}^{-\lambda_1}_9$
   & 3     \\
{ }  \\
odd \hskip 0.3cm &\hskip 0.5cm even\hskip 0.5cm &
     $F^{\lambda_1}_7-\bar{F}^{-\lambda_1}_7$,
     $F^{\lambda_1}_8-\bar{F}^{-\lambda_1}_8$,
     $F^{\lambda_1}_9+\bar{F}^{-\lambda_1}_9$
   & 3     \\
{ }  \\
odd \hskip 0.3cm &\hskip 0.5cm odd \hskip 0.5cm &
     $F^{\lambda_1}_1-\bar{F}^{-\lambda_1}_1$,
     $F^{\lambda_1}_2-\bar{F}^{-\lambda_1}_2$,
     $F^{\lambda_1}_3-\bar{F}^{-\lambda_1}_3$
   & 6     \\
{ }  \\
{ }\hskip 0.3cm &\hskip 0.5cm { }\hskip 0.5cm &
     $F^{\lambda_1}_4+\bar{F}^{-\lambda_1}_4$,
     $F^{\lambda_1}_5-\bar{F}^{-\lambda_1}_5$,
     $F^{\lambda_1}_6-\bar{F}^{-\lambda_1}_6$
    & { }   \\
{ }  \\
\hline
\end{tabular}

\end{enumerate}

\newpage

\section*{Figures}

\begin{enumerate}

\item[{\bf Fig.~1}]
Feynman diagrams which contribute to the process
$\gamma e^-\rightarrow Z e^-$.
The first two diagrams are the SM contribution. The blob
in the last diagram denotes an anomalous vertex.

\item[{\bf Fig.~2}]
Assignments of momenta and helicities for the general
$\gamma ZV$ ($V=\gamma$ or $Z$) vertices.

\item[{\bf Fig.~3}]
A schematic view of the process $\gamma e\rightarrow Ze$.
The indices $\sigma_1$, $\sigma_2$, $\lambda_1$, and $\lambda_2$
denote particle helicities.

\item[{\bf Fig.~4}]
Integrated cross sections ($-0.9<\cos\Theta <0.9$) versus the
$\gamma e$ c.m. energy, $\sqrt{s}$, for left-handed electrons
and various combinations of boson polarizations
$(\lambda_1,\lambda_2)$.

\item[{\bf Fig.~5}]
Differential cross sections for left-handed electrons and
various combinations of boson polarizations $(\lambda_1,\lambda_2)$
at $\sqrt{s}=0.5$ TeV.

\item[{\bf Fig.~6}]
Schematic view of the sequential process
$\gamma e\rightarrow Ze$, $Z\rightarrow f\bar{f}$.
Shown in parentheses are the four-momenta and helicities of
the particles.

\item[{\bf Fig.~7}]
The coordinate system in the colliding $\gamma e$ c.m. frame.
The $y$-axis is chosen parallel to
$\vec{q}_1(\gamma)\times \vec{q}_2(Z)$, and thus it points out of
the page.
The coordinate system in the rest frame is reached from this frame
by a boost along the $z$-axis.

\item[{\bf Fig.~8}]
The average photon helicity of the Compton-backscattered laser light
for the initial unpolarized electron beam and the initial circularly
polarized laser beam ($\lambda_\gamma=1$) at $\sqrt{s}=0.5$ TeV.

\item[{\bf Fig.~9}]
Effective photon spectra of the Compton-backscattered
laser light. The dashed line is for left-handed
photons and the long-dashed line for right-handed
photons.

\item[{\bf Fig.~10}]
Angular distribution ${\rm d}\sigma/{\rm d}\cos\Theta$ at
$\sqrt{s}=0.5$ TeV. Curves are shown for the SM
(solid line), anomalous couplings $h^\gamma_1=0.1$ (long-dashed line)
and $h^Z_1=0.1$ (dot-dashed line). All the other couplings are
as in the SM.

\item[{\bf Fig.~11}]
Allowed regions for the anomalous couplings $h^\gamma_1$ and $h^Z_1$
from $\gamma e\rightarrow Ze$ using backscattered laser beams for
(a) the right-handed initial laser beam and (b) the left-handed
initial laser beam at $\sqrt{s}=0.5$ TeV. All other couplings
assume their SM values.

\item[{\bf Fig.~12}]
Angular dependence of the CP-violation asymmetry,
$A_{CP}(\lambda_\gamma)$, for $\lambda_\gamma=1$ and $h^\gamma_2=0.1$
(solid line), for $\lambda_\gamma=-1$ and $h^\gamma_2=0.1$
(dotted line), for $\lambda_\gamma=+1$ and $h^Z_2=0.1$
(long-dashed line), and for $\lambda_\gamma=-1$ and $h^Z_2=0.1$
(dot-dashed line) at $\sqrt{s}=0.5$ TeV. All other couplings
are as in the SM.
\end{enumerate}
%
%
%
%
%
%
%
%
%
%
%
%
%
%
%
%
%
%
%
%
%
%
%
%
\end{document}